# Superior surface modification layer of poly(styrene) on SiO$_2$ gate insulator in rubrene single crystal field-effect transistor

Hua-Xue Zhou[1], Chong-Li Yang[1], Shi-Lin Li[1], Shota Kobayashi[2], Qifeng Yao[1], Katsumi Tanigaki[1,a)], Hidekazu Shimotani[2,a)]

AFFILIATIONS
[1]Beijing academy of quantum information sciences, Beijing, 100193, China.
[2]Institute of Physics, Chinese Academy of Sciences, Beijing 100190, China. Department of Physics, Graduate School of Science, Tohoku University, 6-3 Aoba, Sendai, Japan.

[a)]Authors to whom correspondence should be addressed: shimotani@tohoku.ac.jp and katsumitanigaki@baqis.ac.cn

**ABSTRACT**
We conduct comparative research on the density of states of electron- and hole- carrier trap levels ($D_{TR}(E)$), dispersing inside the energy gap of a rubrene single crystal in a field effect transistor (FET) struction with Ca and Au hetero-electrodes for an ambipolar carrier injection mode, by using polymeric protection-layer materials on a Si substrate. Three different types of polymeric materials, poly(methyl-methacrylate) (PMMA), poly(styrene) (PS) and poly(chloro-styrene) (PCS) are employed. From the temperature (T)-dependent source-drain current and gate voltage ($I_{SD}$-$V_G$) transfer characteristics, the values of $D_{TR}(E)$ are evaluated. PS exhibits the most efficiently-balanced ambipolar carrier transport, which is superior to PMMA that is most typically used as the standard protection layer on a SiO$_2$/doped-Si substrate. Discussions are made in the framework of a carrier multiple trap and release (CMTR) model.

Organic semiconductors (OSCs) are of particular interest thanks to their light weight, flexibility, printability, low cost, and feasible processing. Important and fascinating characteristics of OSCs are the ambipolar carrier injection and transport of electrons and holes, which are now the most widespread advantages in intrinsic OSCs[1-3]. Different from what is typically observed in inorganic semiconductors (ISCs), efficient and distinctive ambipolar actions of charge carriers have opened an intriguing research area that includes wave-length tuneable and phase coherent lasers in addition to organic light-emitting field effect transistors (OLETs) as a display.

OSCs also provide an excellent platform for fundamental and systematic understanding of the charge transport mechanism in field-effect transistors (FETs)[4,5]. One of the serious characteristic features of OSCs in FETs, however, has been the existing trap levels of injected carriers, which is known to be even more serious in the case of electrons than that of holes[6-8]. Consequently, even in the case of various kinds of electrodes with low work-functions, organic FETs often exhibit p-type electrical conduction but not n-type.

The most typical current benchmark OSC for ambipolar carrier injection and transport is Rubrene (5,6,11,12-tetraphenylnaphthacene, $C_{42}H_{28}$), which exhibits high field-effect mobility[5,9-11] as well as bandlike-transport temperature dependence characteristics[12-14]. The interface protection layer on the gate dielectrics, which makes it possible to reduce the carrier densities captured by the existing trap denisity of states ($D_{TR}(E)$), is important for ambipolar transistor actions. Selection and processing conditions of the interfacial protection layer are crucial for achieving further improvements in FET mobilities[15,16].

In this paper, we study $D_{TR}$ in OSC FET devices based on rubrene single crystals with different polymer thin films as the surface protection layer, by targetting on poly(methylmethacrylate) (PMMA), poly(styrene) (PS) and poly(4-chlorostyrene) (PCS). The distribution of $D_{TR}$ is evaluated based on the temperature (T)-dependence of four-terminal probe FET transport properties. The structure of the protection polymer materials strongly infuences the generation of shallow and deep $D_{TR}$'s both for electrons and holes. It has been demonstrated that all-carbon component PS performs best as protection layer, which is superior to the most frequently used PMMA.

Rubrene single crystals were grown from rubrene powder (Yurui-shanghai chemical Co., Ltd) using physical vapor transport in a three-zone furnace[17]. The typical dimension of the crystal was 3 mm in width, 1 mm in length, and 1 μm in thickness. Single crystal rubrene FETs were fabricated with a PMMA, PC or PCS protection layer on a $SiO_2$/doped-Si substrate (thickness of $SiO_2$: 300 nm, capacitance of $C_{SiO2}$: $6 \times 10^{-9}$ F/cm$^2$) as shown in Fig. 1(a) and (b). Hetero-electrodes consisting of thermally evaporated Ca and Au metals were used for ambipolar carrier injection, which were reported to be efficient for organic FETs previously[3,18] The current–voltage characteristic of the OFETs were measured in a four probe contact configuration by a B1500A semiconductor parameter analyzer (Keysight Technologies, Inc.) at various Ts for analyses. T was controlled by a Peltier device, and the substrate surface T was measured using a platinum resistance thermometer placed on the substrate surface. The measurements were performed in a glove box under an argon gas atmosphere.

The total number (n) of carriers to be accumulated by gate voltage ($V_G$) on a $SiO_2$ dielectric substrate in the FET structure can be expressed by $n=CV_G/e$, where C is the capacitance of a gate insulator and e is the elementary charge of an electron. Given a situation that the majority of accumulated carriers can be accommodated in a trap band existing inside the band gap of either below the conduction band minima (CBM) (ETIB: electron trap impurity band) or above the valence band maxima (VBM) (HTIB: hole trap impurity band) in a small carrier concentration regime, $D_{TR}(E)$ at the energy of E ($E \cong 0$: the energy difference from the intrinsic band) can be calculated by $D_{TR}(E)=(C/e)dV_G(E)/dE$. The position of E existing between CBM and VBM is dependent on $V_G$. In order to evaluate $dV_G(E)/dE$, we employed FET transfer characteristics obtained as a function of T[19,20]. The number of carriers electrically transporting at the interface channel area of an OSC thin film can be expressed by $n_c=n\exp[-E/(k_BT)]$ in the limit of the low carrier concentration regime as seen in Fig. 1(c). The number of the carriers contributing to the electrical conduction ($n_c$) in the channel region can be expressed by $n_c=(CV_G/e)\exp[-E/(k_BT)]$. In this approximation, a simple expression of FET in the linear correspondence is $\sigma_\square = en_c\mu = CV_G\mu\exp[-E/(k_BT)]$, where $\sigma_\square$ is the sheet conductance between the source and the drain, and $\mu$ is the FET device mobility to be evaluated from FET characteristics independently in a conventional manner. The sheet resistance is calculated by $\sigma_\square = LI_{SD}/WV$, where $L$ is the distance between the two central voltage electrodes, $I_{SD}$ is the source-drain current, $W$ is the channel width.

We first explain how we evaluated the DOS(E) by taking example of the PS protection layer. In Fig. 2(a), the $\sigma_\square$–$V_G$ characteristics of a rubrene single crystal FET with a PS protection layer on the SiO$_2$ gate dielectric are displayed for various Ts between 233K to 323K. The activation energy $E$ ($E \gtrsim 0$) from the impurity band to the CBM or the VBM was determined from the T dependent logarismistic $\sigma_\square$-$V_G$ experimental data (ln $\sigma_\square$) by approximating a linear pregression as a function of $1/T$ for each $V_G$. Such a dependence is depicted in Fig. 2(b) at $V_G = 30$ V. When we plot ln $\sigma_\square$ v.s. 1/T at various $V_G$'s for the T-dependent $\sigma_\square$ transfer characteristics, the E($V_G$)-$V_G$ relations can be evaluated as shown in Fig. 2(c) and then d$V_G$(E)/dE was calculated. By employing the derivative, $D_{TR}(E) = (C/e)dV_G(E)/dE$ was quantitatively evaluated. $^eD_{TR}(E)$ corresponding to CB (electrons) was extracted as shown in Fig. 2(d).

The T-dependent $\sigma_\square$-$V_g$ FET transfer characteristics of rubrene FETs fabricated on the three polymers modification layers in the range of T from 230 to 323 K, where $V_{SD}$ =100 V for PS and PCS, $V_{SD}$ =120V for PMMA, are shown in Fig. 3(a) - (c). Ambipolar charge accumulation and transport were observed for the devices of PS and PMMA, while solely p-type charge accumulation and conduction were observed for PCS. Similar systematic evaluations of $D_{TR}(E)$ for both electrons and holes were carried out in Fig. 3(d) - (f). In the case of hole carriers, the $D_{TR}(E)$ distributed near the VBM of rubrene for the all three interface materials, which were shallow impurity levels. Middle and deep carrier trap levels were also observed with smaller intensities for PMMA. However, the situation is greatly different in the case of electron carriers. PS showed the highest intensity $D_{TR}(E)$ near the CBM, while PMMA did not show such a shallow trap level. PMMA showed a trap level in the middle range of energy. These results prove that PS is the best protection interface layer among the three polymeric materials on a SiO$_2$ surface, superior to the most commonly used PMMA.

All the FET configurations are BGTC construction, and therefore there are no significant differences in the number of charge carriers injected to the three types of devices. Consequently, these evaluated $D_{TR}(E)$s emphasize the important information that the charge accumulation is strongly influenced by the interfacial-modification layers in addition to the charge transport in the channel, the latter of which is sensitively modified by the scattering resulting from the interface between rubrene single crystal and the modification layer on a SiO$_2$ substrate. The evaluated $D_{TR}(E)$ values are greatly differentiated depending on the interfacial layer and asymmetric between holes and electrons. This is not due to the change in FET mobility dependence on the interface layer since the mobility was experimentally determined and taken into account for the analysis and corrected in the evaluation process of $D_{TR}(E)$ as follows; The electron mobilities ($^{FET}\mu_e$) were evaluated to be 6.01, 0.75 and 0 cm$^2$/Vs within our experimental resolution, while 5.96, 4.10, and 6.42 cm$^2$/Vs were evaluated for hole mobilities ($^{FET}\mu_e$) for PS, PMMA and PCS, respectively.

The understanding of the $D_{TR}(E)$ inside the gap of OSCs has continuously been under intense debates and various analytical methods are proposed. One model proposes that the $D_{TR}(E)$ originates from the randomly oriented dipoles in the insulating layer near the semiconductor/insulating layer interface[21]. This causes local energy fluctuations and broadening of the trap DOS. Another model is that a carrier at the semiconductor/insulator interface induces ionic polarization of the insulating layer.[22] If the ionic polarization of the insulating layer is strong enough, it will couple with the carrier to form a Fröhlich polaron. When the polaron radius is comparable to the intermolecular distance, carrier transport is proposed to take place by hopping between neighboring molecules. However, it should be kept in mind that such a reduction should not be dissimilar between holes and electrons. Based on what was observed in the present research, the situation was not the case.

In an electric transport process in the framework of the multiple trap and release model (MTRM), the trap band levels play a crucial role for carrier transport of electrons and holes. With the influences of the protection layer, the energy level of electron trap states of organic semiconductors will be deeper as the electron affinity (EA) of the underneath modification layer becomes large. Since the value of EA are homogeneously distributed in the case of polymeric interfacial layer, the distribution of such electron trap states would be broad by large inhomogeneity. The itinerant conduction electrons are trapped in a deeper level in a largely disordered fashion and released to the conduction band in a MTRM process. So, we readily take into account such local EA and ionization energy (IP) factors. The differences in IP of O and Cl constituents from that of C are 2.2 and 1.8eV respectively, and they are comparatively larger than the thermal energy $k_B T$ of 0.026 eV at room T. The differences in EA of the constituents O and Cl are, on the other hand, 0.2 and 2.36 eV, showing a meaningful difference. In the case of holes, the greater IPs of O and Cl, compared to that of C, suggest that such local energy potential does not contribute to the hole transfer in a matrix of carbon atoms, which is different from the situation of electrons. Nonequivalent influence on the carrier type between holes and electrons observed in the present studies seems to be consistent with this understanding.

**Conclusion**

The density of states ($D_{TR}(E)$) of the carrier trap levels inside the gap locating below CBM or above VBM were studied for three types of polymeric interfacial materials, PMMA, PS and PCS, which protect carriers not to be trapped on a $SiO_2$ substrate. $D_{TR}(E)$ values as a function of E were evaluated from T dependent $\sigma$-$V_G$ FET transfer characteristics and discussed in the framework of MTRM model. The $D_{TR}(E)$ sensitively changed depending on the interfacial layers and the transport of electron carriers are even more influenced than that of hole carriers. It proved that all carbon-based PS showed superior balanced-ambipolar carrier transport to that of commonly employed PMMA for the bench-mark organic single crystal rubrene OSC.

**Acknowledgements**
This project has been supported by JSPS KAKENHI Grant Number JP17H05326, JSPS KAKENHI Grant Number JP24684023, JP25610084 and JP16K13826, the National Natural Science Foundation of China (NSFC Grant Nos. 12174027).

**AUTHOR DECLARATIONS**
Conflict of Interest
The authors have no conflicts to disclose.

**Author Contributions**
**Hua-Xue Zhou:** Conceptualization (equal); Formal analysis (equal); Investigation (equal); Methodology (equal); Writing –review & editing (equal); **Chong-Li Yang:** Investigation (supporting); **Shi-Lin Li:** Investigation (supporting); **Shota Kobayashi:** Investigation (equal); Methodology (equal); **Qifeng Yao:** Calculation (lead); **Katsumi Tanigaki:** Conceptualization (equal); Formal analysis (lead); Funding acquisition (lead); Supervision (lead); Writing –review & editing (equal); **Hidekazu Shimotani:** Conceptualization (equal); Investigation (equal); Methodology (equal).

DATA AVAILABILITY
The data that support the findings of this study are available from the corresponding authors upon reasonable request.

Figure and figure captions

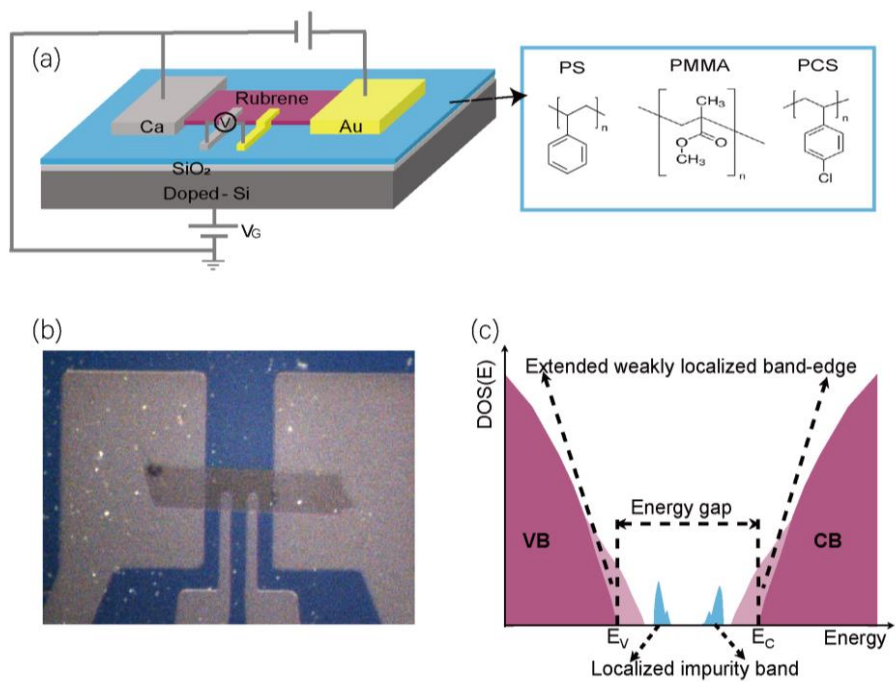

Fig. 1. (a) Schematic diagram of a rubrene based FET with different polymeric materials as a protection layer on a SiO$_2$ gate dielectric of a Si substrate. The molecule structure of PS, PMMA and PCS is shown inside a blue rectangle frame. (b) Photograph of an FET device with BGTE construction. (c) Schematic density of states (DOS), where conduction band (CB) and valence band (VB) (purple color) are shown together with band-edge tails (pale purple in color) connecting to CB or VB and localized impurity band (IB, blue color) in rubrene single crystal.

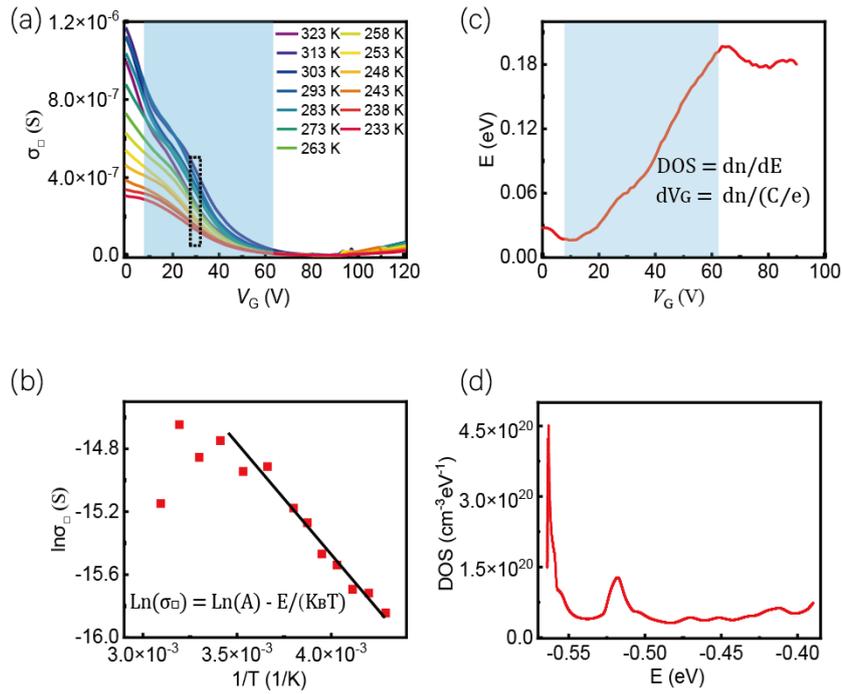

Fig. 2. (a) $\sigma_\square$ - $V_G$ transfer characteristics of rubrene single crystal FET at various temperatures. (b) The activation energy $E$ is extracted from the slope of $\ln \sigma_\square = \ln A - E/(K_B T)$, using the data of T dependent $\ln \sigma_\square$ measured at $V_G = 30$ V shown in the black framework region of (a). (c) For each $V_G$ in (a) at the blue area, the corresponding E is extracted through the linear fitting of $\ln \sigma_\square = \ln A - E/(K_B T)$ and $V_G$ dependent $E$ can be obtained. (d) The DOS of impurity band was evaluated by DOS(= dn/dE) = (C/e)d$V_G$(E)/dE, by employing the data in the blue framework (c).

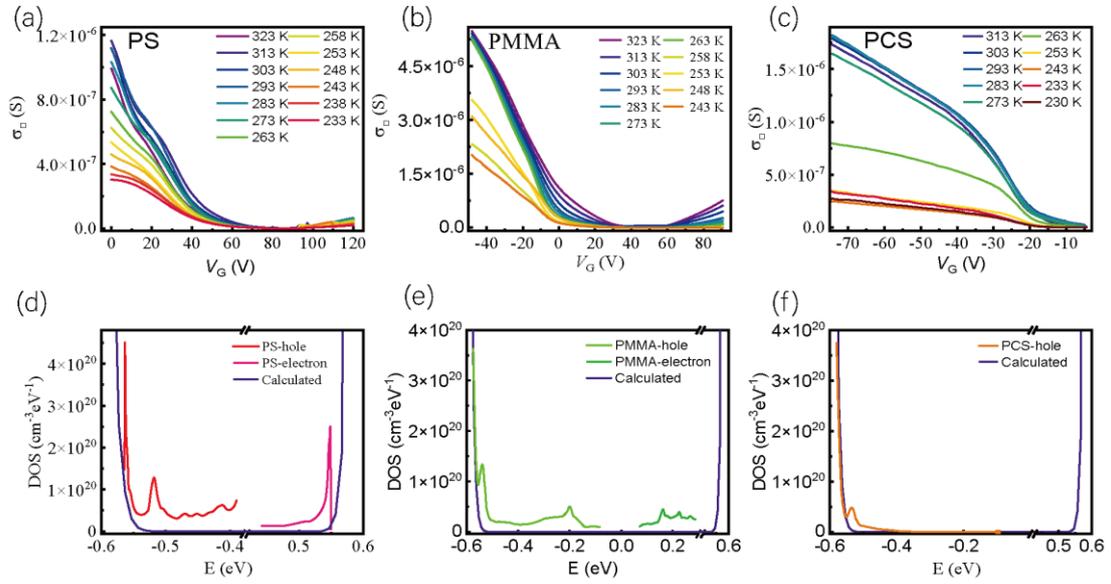

Fig. 3. (a), (b), (c): $\sigma_\square$-$V_G$ transfer curves at various temperatures for the FETs devices with PS, PMMA and PCS protection layers, respectively. (d), (e), (f): Trap DOS distributions as a function of energy E for PS, PMMA and PCS, respectively. The purple line shows CBM and VBM of a rubrene single crystal based on the first principle band caculations. E was scaled as the difference from the band edge as was explained in the main text.